\documentclass[aps, superscriptaddress, showpacs, floatfix, twocolumn]{revtex4-1}

\usepackage{amsmath} \allowdisplaybreaks
\usepackage{hyperref}
\usepackage{graphicx}

\begin{document}
\title{Strain Control of Electronic Phase in Rare Earth Nickelates}
\author{Zhuoran He}
\affiliation{Department of Physics, Columbia University, New York, New York 10027, USA}
\author{Andrew J. Millis}
\affiliation{Department of Physics, Columbia University, New York, New York 10027, USA}

\begin{abstract}
\vspace*{0.5\baselineskip}
We use density functional plus $U$ methods to study the effects of a tensile or compressive substrate strain on the charge-ordered insulating phase of LuNiO$_3$. The numerical results are analysed in terms of a  Landau energy function, with octahedral rotational distortions of the perovskite structure included as a perturbation.  Approximately $4\%$ tensile or compressive strain leads to a first-order transition from an insulating structure with large amplitude breathing mode distortions of the NiO$_6$ octahedra to a metallic state in which breathing mode distortions are absent but Jahn-Teller distortions in which two Ni-O bonds become long and the other four become short are present. Compressive strain produces uniform Jahn-Teller order with the long axis aligned perpendicular to the substrate plane while tensile strain produces a staggered Jahn-Teller order in which the long bond lies in the plane and alternates between two nearly orthogonal in-plane directions forming a checkerboard pattern. In the absence of the breathing mode distortions and octahedral rotations, the tensile strain-induced transition to the staggered Jahn-Teller state would be of second order. 
\end{abstract}

\pacs{68.35.Rh, 71.70.Ej, 72.80.Ga, 73.61.-r}
\maketitle

\section{Introduction\label{sec:intro}}
The rare earth nickelates have been studied for many years  \cite{Torrance92,Alonso99,Khomskii96,Fernandez-diaz00, Staub02} and have been of   substantial recent interest \cite{Chaloupka08,Medarde09,Hansmann09,Ouellette10, Son10, Han11, Benckiser11, Boris11, Scherwitzl11,Chakhalian11,Stewart11a, Stewart11b,Park12,Johnston14,Peil14,Park14} following the proposal of Chaloupka and Khalliulin \cite{Chaloupka08} that in an appropriately chosen superlattice configuration, an electronic structure similar to that found in the high-$T_c$ cuprates could be realized. A one-band state has not been achieved, but the question of the degree to which the electronic structure can be controlled by appropriate combinations of strain and heterostructuring remains an area of active research \cite{Hansmann09,Han11,Benckiser11,Boris11,Park12,Peil14}.   

The chemical formula of the rare earth nickelates is $R$NiO$_3$, with $R$ standing for La or for an element Nd, Pr, Sm, Gd, Eu, Lu of the rare earth series. The materials crystallize in variants of the $AB$O$_3$ perovskite structure with the $R$ ion on the $A$ site and the Ni ion on the $B$ site. The basic structural motif  is a corner-shared $B$O$_6$ octahedron. In the ideal perovskite structure the octahedron has   six equal Ni-O bond lengths and the point group symmetry of the Ni site is O$_h$. The important orbitals are the Ni-centered $e_g$-symmetry $d$ orbitals. Standard valence-counting arguments suggest that the Ni is in the low-spin $d^7$ configuration with a filled $t_{2g}$ shell and one electron in the two $e_g$-symmetry orbitals,  which are degenerate in O$_h$ symmetry.

Having a single electron occupy two degenerate orbitals is expected to favor a symmetry breaking distortion in which one of the $e_g$ orbitals becomes preferentially occupied and the point symmetry of the Ni is lowered from O$_h$ to D$_{4h}$. However, such a distortion has not been observed to date in the nickelate materials. With the exception of the LaNiO$_3$ (which remains undistorted to lowest temperatures), the materials exhibit at low temperatures an ordered phase \cite{Torrance92} characterized by two distinct NiO$_6$ octahedra, one in which the six Ni-O bonds are short (but approximately equal) and one in which the six Ni-O bonds are long (but again approximately equal) \cite{Medarde09}. This disproportionation is sometimes referred to as ``charge ordering'' \cite{Fernandez-diaz00,Staub02} based on the idea that the ionic charge of the Ni with longer Ni-O bond lengths should be larger than of the Ni ions with shorter Ni-O bonds, and based also on a difference in size of measured magnetic moments between the two sites. Although the actual charge difference between the sites is very small \cite{Han11, Park12}, for simplicity we will refer to the disproportionated state as ``charge ordered".

The charge ordering is at first sight surprising  because the dominant interaction in transition metal oxides is generally believed to be a large on-site repulsion ``$U$'' that acts to disfavor charge ordering. Indeed $U=E^{N+1}+E^{N-1}-2E^N$ is defined as the energy cost to change the electronic configuration from $N$-electrons on each transition metal ion to the disproportionated configuration in which half of the ions have $N+1$ electrons and the other half have $N-1$. The behavior is now understood \cite{Khomskii96,Park12} as a consequence of a relatively large electronegativity of Ni. This places the rare earth nickelate materials in or close to the ``negative charge transfer gap'' regime so that the electronic configuration is much closer to $d^8\bar{L}$ than to $d^7$: one electron is transferred from ligand (oxygen) state to Ni so the Ni has two electrons in the  $e_g$ orbitals (in the high-spin configuration)  and there is an average density of $1/3$ hole per O ion. Density functional plus dynamical mean field calculations \cite{Park12,Park14} have shown that in this situation the ``charge disproportionation'' can be understood as a consequence of a hybridization (bond-centered) density wave leading to a site-selective Mott insulating regime. The high-spin $d^8$ configuation of the Ni ions disfavors Jahn-Teller distortions with unequal occupancy of the $e_g$ orbitals in agreement with measurements, indicating that even in strained superlattices of metallic LaNiO$_3$ the difference in occupancy of the two $e_g$ orbitals is small \cite{Benckiser11,Tung13}. Subsequent model system studies confirmed the essential features of this understanding \cite{Johnston14} and suggest that a negative charge-transfer energy implies that the effective low-energy theory is a two-orbital Hubbard-like model with a small or possibly negative effective $U$ but a non-negligible $J$ \cite{Peil14}.

While this physical picture provides a satisfying understanding of the essential features of the observations in terms of specific physics of the nickelate materials, it is incomplete in some respects. First, both experimental and theoretical studies of  orbital disproportionation have focussed on the metallic regimes of the nickelate phase diagram \cite{Han11,Benckiser11,Pesquera12,Wu13,Kinyanjui14} and leaves open the question of strain effects on the physics of the charge-ordered state. Second, the $d^8\bar{L}$ configuration has the same symmetry properties as the $d^7$ configuration (this point was emphasized by Peil et al \cite{Peil14}), meaning that the qualitative arguments suggesting a Jahn-Teller distortion should still apply. If a locally symmetric volume non-preserving disproportionation of the NiO$_6$ octahedra may occur, one may ask why not also cubic-tetragonal disproportionations?

In this paper we address these questions via an \textit{ab initio} study of the response of the charge-ordered insulating ground state of LuNiO$_3$ to an applied biaxial strain. In our study we fully relax the lattice subject to a constraint on the in-plane lattice constant, thereby approximating the effects of substrate-imposed strain on an epitaxially grown film. We find that strains on the order of a few $\%$ (i.e. of a magnitude comparable to those applied by epitaxial growth on reasonable substrate) have the potential to destabilize the charge-ordered state. For compressive strain the result is a metallic state with a modest cubic-to-tetragonal distortion of the NiO$_6$ octahedra, which is moreover approximately the same for each octahedron. Sufficient tensile strain, however, is found to lead to a replacement of charge order by a spatially alternating in-plane Jahn-Teller order. We interpret the calculational results using a Landau theory free energy analysis which provides insights into the orders of the transition (indicating in particular that the transition to staggered Jahn-Teller state is intrinsically of second order and becomes first order only by virtue of competition with the charge-ordered state. Our results thus provide a different perspective on the strain control of orbital properties in transition metal oxides and show that the $d^8\bar{L}$ configuration may also be susceptible to Jahn-Teller order. 

The rest of this paper is organized as follows. Section \ref{Formalism} defines the system we study and the structural distortions we analyse and presents a Landau theory which encapsulates our results. Section \ref{Methods} presents the specifics of our \textit{ab initio} calculations. Section \ref{Results} presents our computational results and Section \ref{Conclusions} is a summary and conclusion.
\\ \ \\

\section{Formalism \label{Formalism}}

\subsection{Structure and strain}

\begin{figure}
\centering
\includegraphics[width=0.9\columnwidth]{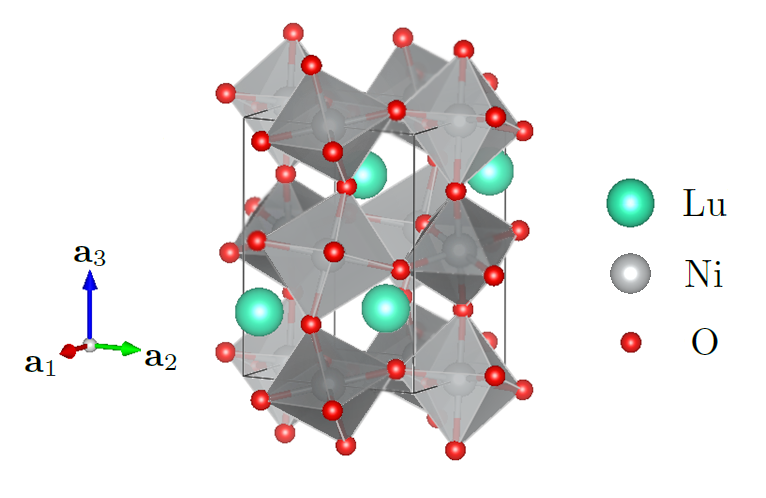}
\caption{``Charge-ordered'' structure of LuNiO$_3$ at vanishing external strain calculated using density functional plus $U$ methods as described in Sec.~\ref{Methods}. NiO$_6$ octahedra are indicated as grey cubes; the darker cubes have mean Ni-O bond length smaller by  $0.10$\AA\ than the lighter ones.   The triad on the left defines the lattice vectors. The calculated lattice constants $|\mathbf{a}_1|=5.12$\AA, $|\mathbf{a}_2|=5.52$\AA, $|\mathbf{a}_3|=7.36$\AA\ are in close agreement with experiment \cite{Alonso01}. We define Cartesian $x$, $y$ and $z$ coordinates so that $z$ is parallel to $\mathbf{a}_3$, and $\mathbf{a}_1$ and $\mathbf{a}_2$ point approximately along the diagonals of the $xy$ plane.  
}
\label{fig:struct}
\end{figure}

In this paper we study LuNiO$_3$. This material has a high charge-ordering transition temperature and an insulating ground state with a large gap to charge excitations \cite{Alonso01}. The ground-state structure obtained from density functional plus $U$ calculations described in Sec.~\ref{Methods} is presented in Fig.~\ref{fig:struct}.  The unit cell has four inequivalent NiO$_6$ octahedra; in the absence of charge ordering, the octahedra differ only by rotations; the charge ordering creates two classes of octahedra with different mean Ni-O bond lengths. Fig.~\ref{fig:struct} also shows the lattice constants. From these we define a Cartesian coordinate system with $z$ axis parallel to $\mathbf{a}_3$ and $x$, $y$ axes in the plane defined by $\mathbf{a}_1$ and $\mathbf{a}_2$ but rotated by $45^\circ$.  

In the ground state of the actual material, the Ni-Ni distance in the basal ($xy$) plane is $3.76$\AA, and there is a slight rhombic distortion, so the Ni-Ni bond angles are $86^\circ$ and $94^\circ$. We wish to simulate the effects of placing LuNiO$_3$ on a substrate, which will typically have a square symmetry. We therefore neglect the  rhombic distortion and consider square structures with $|\mathbf{a}_1|=|\mathbf{a}_2|$ and $90^\circ$ Ni-Ni bond angles in the $xy$ plane. We define the $xy$-plane lattice constant $|\mathbf{a}_1|=|\mathbf{a}_2|=a$ as the diagonal of the square. In the undistorted structure, the lattice constant is $a=a^\star\approx 5.3$\AA, at which the energy is minimum. We will be interested in the consequences of a uniform compression or expansion of the lattice in the $xy$ plane with the $z$ direction free to adjust. We define strain $\delta a$ as an imposed change in the $xy$-plane lattice constant $a$:
\begin{align} \label{eq:straindef}
\delta a=a-a^\star.
\end{align}

The key variables in response to $\delta a$ are the shapes, sizes and orientations of the NiO$_6$ octahedra, which are the important dynamical variables of the structure. In the rest of this section we build up a theoretical description of the relevant distortions, starting from the simple case of an isolated NiO$_6$ octahedron and adding complexity as needed. 

\subsection{An isolated NiO$_6$ octahedron}

To define notation we begin by consideration of one isolated NiO$_6$ octahedron. The unstrained structure is perfectly cubic (point symmetry O$_h$) with 6 mutually perpendicular Ni-O bonds, which we take to lie in the $\pm x$, $\pm y$ and $\pm z$ directions. All the 6 bonds have the same length $l_0\approx 2$\AA.

The distortions of interest here preserve the inversion symmetry about the Ni and the orthogonality of the Ni-O bonds, so that a D$_{2h}$ symmetry is preserved. The distortions may be expressed in terms of three modes, defined in terms of the changes $\delta l_x$, $\delta l_y$, $\delta l_z$ in the $x$, $y$, and $z$ bond lengths as
\begin{subequations}
\begin{align}
Q_0&=\frac{\delta l_x+\delta l_y+\delta l_z}{\sqrt{3}},\\
Q_1&=\frac{\delta l_x-\delta l_y}{\sqrt{2}},\\
Q_3&=\frac{-\delta l_x-\delta l_y+2\delta l_z}{\sqrt{6}}.
\end{align}
\end{subequations}
Here $Q_0$ is the volume expansion mode, $Q_1$ the (volume-preserving) $xy$-plane square-to-rhombic distortion, and $Q_3$ the (volume-preserving) cubic-to-tetragonal Jahn-Teller distortion in $z$ direction. The $Q_0$ mode is invariant under O$_h$. The $Q_1$ and $Q_3$ modes together form a two-dimensional irreducible representation. Therefore, the energy of the octahedron $E(Q_0,Q_1,Q_3)$ to cubic order will be of the form
\begin{align} \label{eq:E-isolated}
E=AQ_0^2&+B(Q_1^2+Q_3^2)\nonumber\\
&+C\left(Q_1^2-\frac{Q_3^2}{3}\right)Q_3+\cdots,
\end{align}
where $A$, $B$ and $C$ are constants. We omitted the cubic terms $Q_0^3$ and $Q_0(Q_1^2+Q_3^2)$, which are just products of lower-order O$_h$ invariants, and highlight the cubic coupling $Q_1^2Q_3$ in the third term with coefficient $C$. In the lattice system, this part will give rise to an important coupling between the distortion $Q_3^\mathbf{F}$ and the staggered Jahn-Teller order $Q_1^\mathbf{C}$, which we will define later.

\subsection{A corner-shared NiO$_6$ array \label{sec:oct-array}}

We next consider an infinite 3D crystal of NiO$_6$ octahedra, still with the O$_h$ symmetry in the unstrained structure at each Ni site. We must now attach a momentum label to each mode. In addition, because the octahedra are corner-shared, there are constraints on the allowed momenta for each distortion. The momenta of interest are $\mathbf{F}=(0,0,0)$, $\mathbf{G}=(\pi,\pi,\pi)$, $\mathbf{C}=(\pi,\pi,0)$. Note that these momenta are defined in the unit cell of the ideal cubic structure with one octahedron per unit cell. Of primary interest in interpreting the numerical results are the modes
\begin{subequations}
\begin{align}
q_0&=Q_0^\mathbf{G};\quad \mbox{two-sublattice ``charge order''},
\label{eq:q0def}
\\
q_1&=Q_1^\mathbf{C};\quad
\mbox{in-plane staggered Jahn-Teller}.
\label{eq:q1def}
\end{align}
In addition, it will be useful to consider
\begin{align}
Q_0&=Q_0^\mathbf{F};\quad
\mbox{volume change},
\label{eq:Q0def}
\\
Q_3&=Q_3^\mathbf{F};\quad
\mbox{uniform Jahn-Teller},
\label{eq:Q3def}
\\
q_3&=Q_3^\mathbf{G};\quad
\mbox{two-sublattice Jahn-Teller},\qquad\,
\label{eq:q3def}
\end{align}
\end{subequations}
which describe the response to a uniform strain and its coupling to a two-sublattice charge order.

The energy function $E(Q_0,Q_3,q_0,q_1,q_3)$ of the 5 modes is in general very complicated. A group theoretical analysis is given in Appendix \ref{appendix:A}. The variables $Q_0$ and $Q_3$ are controlled by the strain $\delta a$, which induces a $Q_3$ distortion and, via Poisson-ratio considerations, a nonzero volume change $Q_0$ of opposite sign to $Q_3$. Both $Q_0$ and $Q_3$ are coupled to the order parameters $q_0$, $q_1$ and $q_3$, and these couplings will drive the phase transitions of interest. Our numerical results to be presented in Sec.~\ref{Results} may be understood in terms of an energy function $E(q_0,q_1|\delta a)$ involving $q_0$ and $q_1$ only, with the other variables $Q_0$, $Q_3$ and $q_3$ determined by the strain $\delta a$ and the values of $q_0$ and $q_1$.

Next, we focus on $q_0$, the two-sublattice charge order. Viewed as a function of $q_0$ only (i.e. when $q_1=0$), our results indicate that the energy $E$ has a first-order transition structure
\begin{align} \label{eq:Eq0}
E(q_0)=A_{20}q_0^2+A_{40}q_0^4+A_{60}q_0^6,
\end{align}
with strain-dependent coefficients $A_{20}, A_{60}>0$ and $A_{40}<0$ near the transition. Thus, $E(q_0)$ has three local minima, at $q_0=0$ and $q_0=\pm q^\star$. The strain $\delta a$ turns out to affect the value of $q^\star$ only slightly; the main effect is to control via $Q_0$ and $Q_3$ the energy difference $\Delta E=E(0)-E(\pm q^\star)$, which is plotted against the in-plane lattice constant $a$ in Fig.~\ref{fig:delta-E}.

We next consider the energy as a function of $q_1$ only (i.e. $q_0=0$). This energy will be found to have a second-order transition structure
\begin{align} \label{eq:Eq1}
E(q_1)=A_{02}q_1^2+A_{04}q_1^4,
\end{align}
with (at zero strain) $A_{02},A_{04}>0$. Applying a strain $\delta a$ leads to nonzero $Q_0$ and $Q_3$ (see Fig.~\ref{fig:modes}(a)). Both of these may couple linearly to $q_1^2$ (recall in particular the $Q_3q_1^2$ cubic invariant), so that we have
\begin{align} \label{eq:A02}
A_{02}=A_{02}^{(0)}-A_{02}^{(1)}\delta a,
\end{align}
indicating that at tensile strains $\delta a>A_{02}^{(0)}/A_{02}^{(1)}$, an in-plane staggered Jahn-Teller order could be favored.

Finally, there is a biquadratic coupling $A_{22}q_0^2q_1^2$ between the charge and Jahn-Teller orders; the sign of $A_{22}>0$ is such that the two orders compete with each other. The resultant energy is therefore
\begin{align} \label{eq:Eq0q1}
E_{\mathrm{cubic}}(q_0,q_1)=E(q_0)+E(q_1)+A_{22}q_0^2q_1^2.
\end{align}

\subsection{Including octahedral rotations}

With Eq.~\eqref{eq:Eq0q1} in hand, we now consider the structure of the actual materials. This involves a GdFeO$_3$-type rotational distortion with four inequivalent Ni ions (see Fig.~\ref{fig:struct}). The O$_6$ octahedron around a given Ni site is rotated; the rotations may be symbolically written as $\alpha_z^+ \beta_x^- \beta_y^-$, meaning that starting from the ideal cubic perovskite structure there is a rotation by angle $\alpha$ about the $z$ axis, and by angle $\beta$ about the $x$ and $y$ axes. The superscript ``$+$'' means the $\alpha$ rotations in neighboring octahedra about the rotational axis of $\alpha$ (the $z$ axis) are in the same direction, while the ``$-$'' means the $\beta$ rotations in neighboring octahedra about the rotational axis of $\beta$ ($x$ or $y$ axis) are in opposite directions. The angles $\alpha$ and $\beta$ are small enough ($<15^{\circ}$ in LuNiO$_3$) that we may neglect the non-Abelian aspect of rotations and treat them as commuting (additive) axial vectors, rather than non-commuting (multiplicative) second-rank tensors.

The important feature of the octahedral rotations is a breaking of the $q_1\leftrightarrow-q_1$ symmetry while preserving the $q_0 \leftrightarrow -q_0$ symmetry. Here the k-points $\mathbf{G}=(\pi,\pi,\pi)$ of $q_0$ and $\mathbf{C}=(\pi,\pi,0)$ of $q_1$ are defined with respect to the undistorted structure, i.e. with one octahedron per unit cell. This is allowed because even though the distorted and rotated structure now has 4 translationally inequivalent Ni ions, the Landau energy function of the system is still invariant under any translation by a nearest-neighbor Ni-Ni distance.

The octahedral rotations generate an energy term that is linearly proportional to $q_1$, and is of order $\alpha \beta^2 \simeq 10^{-2}$ in radians. The derivation of this is in Appendix \ref{appendix:B} using group theory again. Similarly, a term $Q_0q_1$ or $Q_3q_1$ becomes allowed in addition to the $Q_3q_1^2$ term that we previously discussed. Thus, the final energy function is given by
\begin{align} \label{eq:Efinal}
E&=E_{\mathrm{cubic}}-A_{01}q_1\nonumber\\
&=E_{\mathrm{cubic}}-(A_{01}^{(0)}+A_{01}^{(1)}\delta a)q_1,
\end{align}
where $A_{01}^{(0)}$ and $A_{01}^{(1)}$ are by a factor of $\alpha \beta^2 \simeq 10^{-2}$ smaller than the coefficients in $E_{\mathrm{cubic}}$, and $A_{01}^{(1)}$ results from the coupling terms $Q_0q_1$ and $Q_3q_1$.
The linear term is found to change sign at a compressive strain $\delta a=-A_{01}^{(0)}/A_{01}^{(1)}$. We will look into the details in Sec.~\ref{Results}. The added term $-A_{01}q_1$ has an effect similar to that of an external magnetic field on a system near a ferromagnetic transition.

Eq.~\eqref{eq:Efinal} provides a minimal model that explains our data in all important qualitative aspects. In reality, there can be higher order terms of $q_0$ and $q_1$ in both $E_{\mathrm{cubic}}$ and the symmetry breaking terms, as well as nonlinear dependence of the coefficients $A_{nm}$ on strain $\delta a$.

\section{Methods \label{Methods}}

In Sec.~\ref{Formalism}, we constructed a Landau energy function of the bond-length distortion modes $q_0$ and $q_1$ in $R$NiO$_3$ and took into account the GdFeO$_3$-type rotational distortions $\alpha_z^+\beta_x^-\beta_y^-$ of the NiO$_6$ octahedra in a perturbative way. In this section and Sec.~\ref{Results}, we study the structural transitions of LuNiO$_3$  numerically by doing  \textit{ab initio} density functional$\:+\:U$ (DFT+$U$) calculations including structural relaxation. The results are then interpreted using the Landau energy function in Eq.~\eqref{eq:Efinal}.

Our calculations use the Vienna \textit{ab initio} simulation package (VASP) \cite{Kresse96}. The LDA+$U$ algorithm we choose in VASP is the rotationally invariant LSDA+$U$ that follows Ref. \cite{Liechtenstein95}. The Hubbard $U$ of the Ni $3d$ orbitals in LuNiO$_3$ can be obtained by various methods, e.g. constrained LDA \cite{Gunnarsson89,Gunnarsson90}, self-consistent linear response \cite{Cococcioni05}, constrained RPA \cite{Springer98,Kotani00}, etc. They all give values of $U$ within $U=(5\pm 1)$ eV. The Hund's coupling $J$ is estimated to be $0.5\sim 1$ eV. We finally chose $U=5$ eV and $J=1$ eV, as they gave a structure in Fig.~\ref{fig:struct} that was closest to the experimental results. Slight changes of $U$ and $J$ within their errors were tried and no qualitative difference was found.

We did a spin-polarized calculation using the PAW-PBE pseudopotential provided by VASP. The k-point mesh we used was $6\times 6\times 6$ and the cut-off energy of the plane-wave basis was set to $600$~eV. The errors due to k-points and energy cut-off are estimated to be smaller than the errors due to $U$ and $J$ by comparing with results obtained using a coarser k-point mesh of $4\times 4\times 4$ or a lower energy cut-off of $400$~eV.

The computational unit cell was chosen to contain four LuNiO$_3$ formula units. Defining the basal plane as the one in which strain is applied, we take two formula units in the basal plane and two displaced vertically.  To mimic the effects of a substrate, the in-plane lattice constants $|\mathbf{a}_1|=|\mathbf{a}_2|=a$ are fixed to pre-set and equal values (so any in-plane rhombic distortion is neglected). $|\mathbf{a}_3|$ and all of the intra-unit cell degrees of freedom are allowed to relax. We slightly modified the conjugate gradient code in VASP to do this.

The minimum energy of the substrate-constrained system is obtained at $a=a^\star\approx 5.3\mbox{\AA}$. The structure obtained is almost identical to the free structure in Fig.~\ref{fig:struct}, except that $|\mathbf{a}_1|$ and $|\mathbf{a}_2|$ are made equal (the small rhombic distortion is suppressed). We then adjust the substrate lattice constant $a$, our control parameter, away from $a^\star$ and see how the structure changes.

\section{Results \label{Results}}

\subsection{Structures and energy difference}

\begin{figure}
	\centering
	\includegraphics[width=0.9\columnwidth]{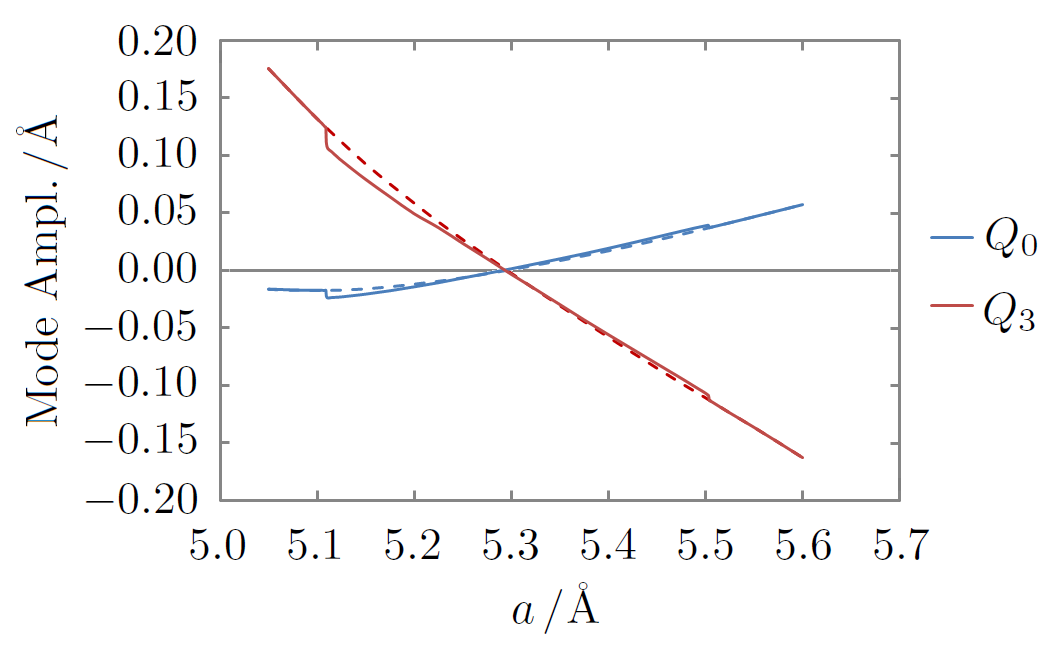}\\
	\centering (a)\\
	\includegraphics[width=0.9\columnwidth]{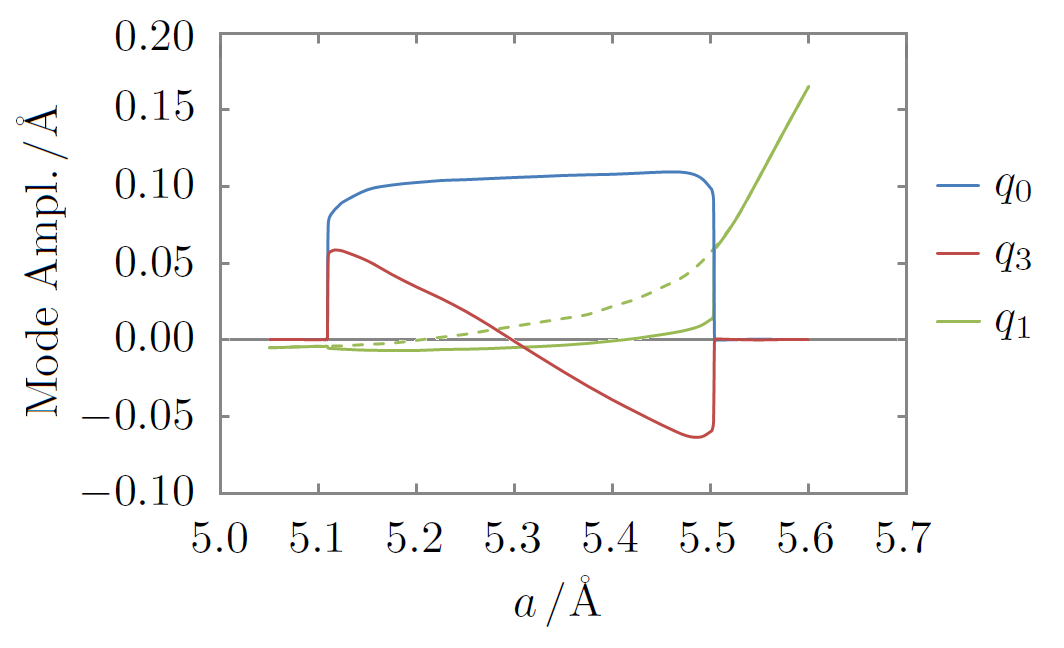}\\
	\centering (b)\\
	\caption{Upper panel: strain dependence of spatially uniform volume-changing ($Q_0$) and even-parity volume-preserving cubic-tetragonal ($Q_3$) octahedral modes. Lower panel: strain dependence of staggered volume-changing ($q_0$) and two different even-parity volume-preserving cubic-tetragonal octahedral modes ($q_1$ and $q_3$). Solid lines: results obtained from energy minimization. Dashed lines: results obtained from metastable states obtained by forcing staggered charge order ($q_0$) modes to zero.}
	\label{fig:modes}
\end{figure}

Fig.~\ref{fig:modes} presents our main computational results: the evolution with strain of the structural parameters defined in Sec.~\ref{sec:oct-array}. The upper panel shows the spatially uniform component of the relevant distortions. We see that an applied strain, as expected, induces both a uniform distortion of the NiO$_6$ octahedra (the $Q_3$ mode) and a volume change. The changes are approximately linear in the applied strain. The lower panel shows that  in the absence of strain the ground state is charge ordered ($q_0\neq 0$). Modest strain does not change the amplitude of the charge order but does activate a modest amplitude of staggered Jahn-Teller order ($q_3$) as expected from the combination of strain and breaking of translational symmetry. When the strain exceeds a critical value (in either the compressive or the tensile direction) the charge order vanishes via a first-order transition. On the compressive strain side, the resulting state is metallic (as seen from the value of the density of states at the Fermi surface, not shown), and characterized by no order except that imposed by the strain. On the tensile strain side, the staggered charge order is replaced by a staggered Jahn-Teller order ($q_1$) of comparable magnitude to the charge order. The Jahn-Teller order $q_1$, unlike the charge order $q_0$, does not open a gap at the Fermi level, resulting in a metallic state. In this state, the long-bond direction of the Jahn-Teller order lies  in plane. We also see that a very small-amplitude version of this order exists even in the charge-ordered phase, and the dashed lines show that if the charge order is suppressed, the amplitude of the Jahn-Teller order $q_1$ dramatically increases.

\begin{figure}
\centering
\includegraphics[width=0.85\columnwidth]{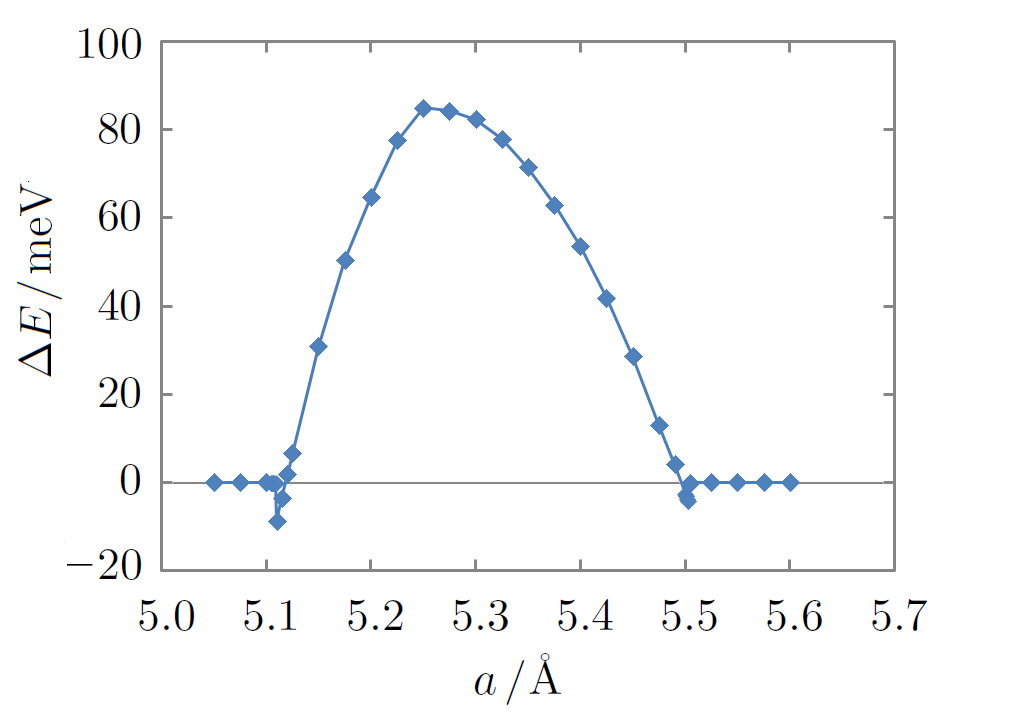}
\caption{The energy difference $\Delta E=E_{\mathrm{JT}}-E_{\mathrm{CO}}$ at different lattice constants $a$, with $E_{\mathrm{JT}}$ and $E_{\mathrm{CO}}$ denoting the energies of the metastable Jahn-Teller distorted structure (dashed lines in Fig.~\ref{fig:modes}) and the stable charge-ordered structure (solid lines in Fig.~\ref{fig:modes}) between the transition points $a\approx 5.1$\AA\ and $a\approx 5.5$\AA . Outside the transition points $\Delta E=0$ because the charge-ordered structure does not exist and relaxes to the only stable Jahn-Teller structure.}
\label{fig:delta-E}
\end{figure}

The energy difference between the metastable and stable states in Fig.~\ref{fig:modes} is plotted in Fig.~\ref{fig:delta-E}. At zero strain $a=a^\star\approx 5.3$\AA , the charge-ordered (CO) structure is lower in energy than the Jahn-Teller (JT) structure by $82$~meV per computational unit cell as defined in Fig.~\ref{fig:struct}. Under either a compressive strain ($a<a^\star$) or a tensile strain ($a>a^\star$), the Jahn-Teller structure is favored and $\Delta E$ is reduced. At both transition points, the curve overshoots a little bit to below zero, and ends at where the charge-ordered structure becomes locally unstable and relaxes to the Jahn-Teller structure. Both the overshoot and the linear $\Delta E-a$ relation near the transitions confirm that the transitions are first order.

\subsection{The lower transition}

\begin{figure}
\centering
\includegraphics[width=0.95\columnwidth]{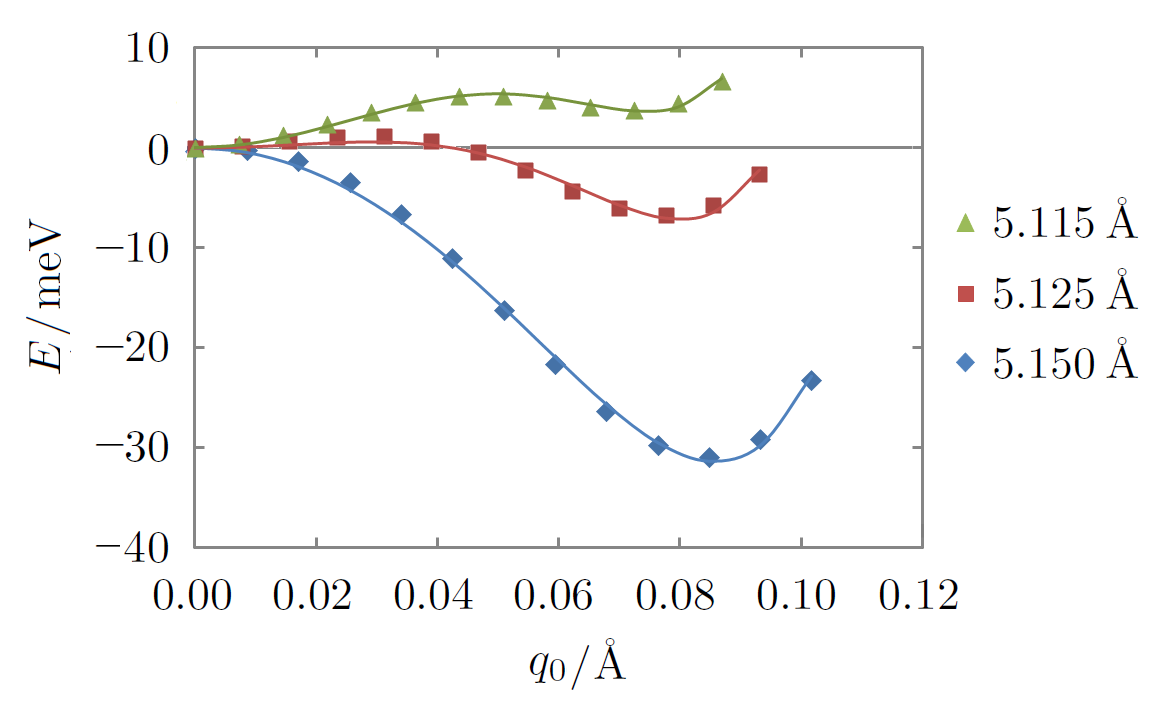}
\caption{The energy plots of linearly interpolated structures between the Jahn-Teller ($q_0=0$) and charge-ordered (minimum at $q_0=q^\star$) states under compressive strains. The energy of the Jahn-Teller structure with $q_0=0$ is used as a reference point and the energies of other structures are measured relative to it. The data points are fitted to Eq.~\eqref{eq:Eq0}, with $A_{60}>0$ for all three curves. The other coefficients satisfy $A_{20}>0, A_{40}<0$ for $a=5.115$\AA\ and $a=5.125$\AA , and $A_{20}<0, A_{40}>0$ for  $a=5.150$\AA .}
\label{fig:E-interp}
\end{figure}

In this subsection we analyse the compressive strain-driven transition with the help of the Landau energy function in Eq.~\eqref{eq:Efinal}. At fixed lattice constant $a$, we calculated the energy of a series of structures linearly interpolated between the structure with $q_0=0$ and the charge-ordered ground-state structure. Results are plotted in  Fig.~\ref{fig:E-interp}. We see that the energy has the typical first-order structure, with two locally stable minima crossing in energy as the lattice constant $a$ is varied and we also see that the value of $q_0$ characteristic of the charge order minimum is insensitive to the value of the strain. The main effect is simply a coupling of the strain to the energy difference.

The Landau energy function in Eq.~\eqref{eq:Efinal} is reduced to Eq.~\eqref{eq:Eq0} near the lower transition point $a\approx 5.1$\AA\ , because $q_1\approx 0$ (see Fig.~\ref{fig:modes}(b)) makes $q_0$ the only order parameter to consider. All coefficients $A_{20}$, $A_{40}$ and $A_{60}$ are found to simultaneously change with the strain $\delta a$.

The transition can nevertheless be understood by a strain-induced change in $A_{20}$ alone. At zero strain $a=a^\star\approx 5.3$\AA\ (not plotted), the coefficient $A_{20}$ is negative, so that the ground-state structure is charge ordered. A compression of the in-plane lattice constant $a$ favors the stability of the out-of-plane Jahn-Teller structure with $q_0=0$, thereby increasing the quadratic coefficient $A_{20}$. Sure enough, we see $A_{20}$ change sign in Fig.~\ref{fig:E-interp} as $a$ decreases from $5.150$\AA\ to $5.125$\AA . Then $q_0=0$ becomes lower in energy than $q_0=q^\star$, the charge order minimum, when $a$ is further reduced. And finally, at some $a$ below $5.115$\AA , the local minimum $q=q^\star$ disappears and $q_0=0$ becomes the only equilibrium structure.

The transition from a nonzero $q_0=q^\star$ to $0$ can be either first order or second order, depending on the sign of the quartic coefficient $A_{40}$ near the transition point $a\approx 5.1$\AA . Since our data shows a first-order transition, we know $A_{40}<0$ and the charge-ordered phase with $q_0=q^\star$ is stabilized by $A_{60}>0$, i.e. the sixth-order term. This agrees with the fit parameters in Fig.~\ref{fig:E-interp}.

\subsection{Evolution of the Jahn-Teller structure}

\begin{figure}
\centering
\includegraphics[width=\columnwidth]{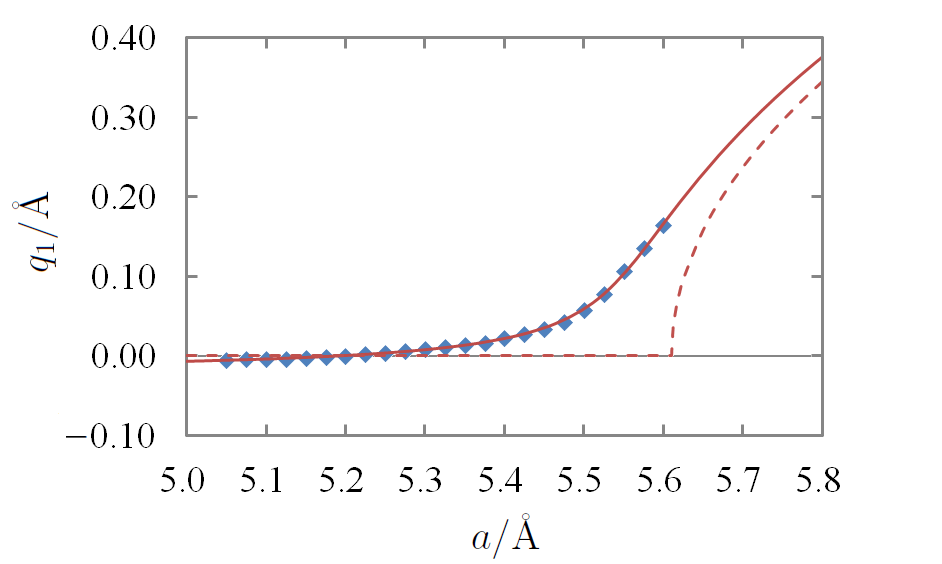}
\caption{Dependence of amplitude $q_1$ of staggered in-plane Jahn-Teller distortions on applied strain. Points are calculated values. Solid line is result of fitting calculated points to  Eq.~\eqref{eq:daq1}.  The solid line is the best-fit line and the dashed line is obtained by setting the linear coefficients $A_{01}^{(0)}=A_{01}^{(1)}=0$ in Eq.~\eqref{eq:daq1} to recover the ideal case of a second-order phase transition. The parameters of the best-fit line are $A_{01}^{(0)}=5.89\times 10^{-3}$, $A_{01}^{(1)}=5.61\times 10^{-2}$, $A_{02}^{(0)}=0.388$, $A_{02}^{(1)}=1.253$, $A_{04}=1$, and $a^*=5.30$\AA .}
\label{fig:q1-a_plot}
\end{figure}

The higher transition at $a\approx 5.5$\AA\ is more complicated because it involves both $q_0$, the breathing mode, and $q_1$, the in-plane staggered Jahn-Teller mode, together with the GdFeO$_3$-type octahedral rotations $\alpha_z^+\beta_x^-\beta_y^-$ that break the $q_1\leftrightarrow-q_1$ symmetry. We begin our analysis of this transition by considering calculations in which $q_0$ is artificially set to zero. The heavy points in Fig.~\ref{fig:q1-a_plot} (equivalent to the dashed $q_1$-line in Fig.~\ref{fig:modes}(b)) show the calculated evolution of $q_1$ with strain when $q_0=0$. We see that over the whole range $q_1\neq 0$, and that the evolution with strain is nonlinear.  The nonzero $q_1$ is a consequence of the GdFeO$_3$ rotations, which, as previously discussed, couple linearly to the staggered component of the Jahn-Teller distortion.

A minimal model to understand this evolution of the Jahn-Teller structure can be obtained by setting $q_0=0$ in  Eqs.~\eqref{eq:Eq1}--\eqref{eq:Efinal}, leading to 
\begin{align} \label{eq:Eq1-final}
E(q_1)=&-(A_{01}^{(0)}+A_{01}^{(1)}\delta a)q_1
\nonumber\\
&+(A_{02}^{(0)}-A_{02}^{(1)}\delta a)q_1^2+A_{04}q_1^4,
\end{align}
where $A_{04}$ is assumed constant for simplicity.  Eq.~\eqref{eq:Eq1-final}  is formally similar to the equation describing a ferromagnet in a magnetic field. The coefficients $A_{01}^{(0)}$ and $A_{01}^{(1)}$ are like an external magnetic field in the ferromagnetic case and arise from the breaking of $q_1\leftrightarrow -q_1$ symmetry due to the GdFeO$_3$ rotations. The need to allow for a strain dependence of the coefficients is shown by the zero crossing of $q_1$ at $a=a_1=5.20$\AA. The dependence of $A_{02}$ on strain reflects the tendency of tensile strain to favor the staggered Jahn-Teller order $q_1$.

Minimizing Eq.~\eqref{eq:Eq1-final} leads to 
\begin{align} \label{eq:daq1}
\frac{-A_{01}^{(0)}+2A_{02}^{(0)}q_1+4A_{04}q_1^3}{A_{01}^{(1)}+2A_{02}^{(1)}q_1}=\delta a.
\end{align}

We have fit Eq.~\eqref{eq:daq1} to the data points shown in Fig.~\ref{fig:q1-a_plot} and from the fit parameters we extracted the critical lattice constant $a=a_2=5.61(4)$\AA\ at which the hypothetical cubic structure would be unstable to staggered Jahn-Teller order, in the absence of charge order or GdFeO$_3$ rotations. We observe that while the uncertainties involved in fitting a four-parameter function to the data mean that individual coefficients cannot be determined with high accuracy, the estimated $a_2$ is robust.  It is interesting that this value is not very much larger than $5.5$\AA\ at which the charge order vanishes. 

\subsection{The competition between $q_0$ and $q_1$}

Comparison of the solid and dashed lines in Fig.~\ref{fig:modes} shows that the staggered charge order ($q_0$) strongly suppresses the staggered Jahn-Teller order ($q_1$). In the notation of Eq.~\eqref{eq:Eq0q1}, the biquadratic term $A_{22}q_0^2q_1^2$ is large and repulsive. In terms of the analysis of Eq.~\eqref{eq:Eq1-final}, $A_{02}$ becomes $A_{02}+A_{22}q_0^2$ and is so much more positive that until the charge order collapses at a first-order transition the staggered Jahn-Teller order cannot develop. There is therefore a strong competition between the two staggered orders $q_0$ and $q_1$.

\section {Conclusions \label{Conclusions}}

We have used density functional and Landau theory methods to consider the effect of strain (induced by growth on a substrate with different lattice constants) on the charge-ordered state of LuNiO$_3$. We find that the charge-ordered state plays a primary role in controlling the physics. It is the leading instability under ambient conditions, and its presence suppresses any other instabilities. However, with sufficient applied strain (within the DFT+$U$ approximation, of the order of $4\%$) the system undergoes a first order transition to a non-charge-ordered state. Interestingly, for tensile strain, the non-charge-ordered state is characterized by a staggered Jahn-Teller order. 

In the actual crystals, the symmetry breaking induced by the GdFeO$_3$ rotational distortion means that the staggered Jahn-Teller order does not break any additional symmetry of the system, but our Landau theory analysis indicates that even the ideal cubic nickelate will undergo a transition to staggered Jahn-Teller order if the tensile strain amplitude is sufficiently large.  We thus conclude that even  a $d^8{\bar L}$ system with a negative charge-transfer energy may have a Jahn-Teller instability. 

The actual magnitude of the strain needed to destabilize the charge order and allow other states is an important open question. While we imagine the strain as being produced by epitaxial growth on a substrate we have not included any quantum confinement effects in our modelling. Also, the DFT+$U$ method we have used is known to overestimate the tendency to charge order \cite{Park14}. The charge order phase boundary also depends on how the double counting correction is implemented.   More refined calculations, perhaps based on DFT+DMFT methods, should be employed to obtain better estimates for the strain needed to destabilize the charge order. But it is interesting that the magnitude of strain we have found is of the order of strains accessible by epitaxial growth on substrates. On the other hand, the first order nature of the transition means that there are no significant precursor effects to the transition.

\textit{Acknowledgements:} We thank C. Marianetti, H. Park and H. Chen for helpful discussions and the Basic Energy Sciences Program of the US Department of Energy for support under grant DOE-ER-046169. 

\bibliography{strain_control_refs}

\begin{widetext}
\newpage
\end{widetext}

\section*{Appendix}
\numberwithin{equation}{subsection}
\makeatletter
\renewcommand{\p@subsection}{}
\makeatother

\subsection{The Landau energy function based\\ on cubic (O$_h$) symmetry \label{appendix:A}}

In the Fourier space, the nonzero distortion modes that appear in the calculated structures of LuNiO$_3$ (other rare-earth nickelates should have very similar perovskite structures) are $Q_0^\mathbf{F}$, $Q_3^\mathbf{F}$ at k-point $\mathbf{F}=(0,0,0)$, which are the uniform expansion and cubic-to-tetragonal distortions of all octahedra, $Q_1^\mathbf{C}$ at k-point $\mathbf{C}=(\pi,\pi,0)$, which characterizes a checkerboard pattern of in-plane staggered Jahn-Teller distortions, and $Q_0^\mathbf{G}$, $Q_3^\mathbf{G}$ at k-point $\mathbf{G}=(\pi,\pi,\pi)$, which appear only in the ``charge-ordered'' structure.

To find out the symmetry-determined form of the Landau energy $E$ as a function of the 5 modes $Q_0^{000}$, $Q_3^{000}$, $Q_1^{\pi\pi 0}$, $Q_0^{\pi\pi\pi}$, and $Q_3^{\pi\pi\pi}$, we need to extend our configuration space to a minimal O$_h$ group-invariant subspace of 9 dimensions. This is because the Jahn-Teller distortion $Q_3^{000}$ along the $z$ direction can be rotated to $x$ and $y$ directions by O$_h$ to give us the $Q_1^{000}$ mode. Similarly, rotating $Q_3^{\pi\pi\pi}$ to $x$ and $y$ directions gives us $Q_1^{\pi\pi\pi}$. And $Q_1^{\pi\pi 0}=\delta l_x^{\pi\pi 0}-\delta l_y^{\pi\pi 0}$ can be rotated to $Q_1^{0\pi\pi}=\delta l_y^{0\pi\pi}-\delta l_z^{0\pi\pi}$ and $Q_1^{\pi 0\pi}=\delta l_z^{\pi 0\pi}-\delta l_x^{\pi 0\pi}$ and, of course, mirrored to $-Q_1^{\pi\pi 0}$, $-Q_1^{0\pi\pi}$, and $-Q_1^{\pi 0\pi}$. We now see the 9 orthonormal modes
\begin{align} \label{eq:9-modes}
\begin{array}{lll}
Q_0^{000},& Q_1^{000},& Q_3^{000},\\
Q_0^{\pi\pi\pi},& Q_1^{\pi\pi\pi},& Q_3^{\pi\pi\pi},\\
Q_1^{\pi\pi 0},& Q_1^{0\pi\pi},& Q_1^{\pi 0\pi},
\end{array}
\end{align}
as a basis of the 9-dimensional extended configuration space that is invariant under O$_h$. 

The Landau energy $E$ as a function of the 9 modes in Eq.~\eqref{eq:9-modes} will have to be invariant under the $3!=6$ permutations of the $x$, $y$, and $z$ indices due to O$_h$, and the translations along $x$, $y$ and $z$ as well. A translation along $x$ by one nearest neighbor Ni-Ni distance, for example, will leave all $k_x=0$ modes unchanged and let all $k_x=\pi$ modes change sign. Translations in all 3 directions can generate totally $2^3=8$ ways of sign change. The Landau function $E$ will therefore have to be invariant under $6\times 8=48$ symmetry operations which include O$_h+\mbox{translations}$. At this point we forget about the boundary effects of the thin film and the substrate, so that the $x$, $y$, and $z$ directions are all equivalent in the extended 9-dimensional configuration space.

The algorithm we use for determining the symmetry-allowed form of the energy $E$ is mainly based on the rearrangement theorem of group theory. We start with a general Taylor expansion of $E$ with respect to the 9 variables in Eq.~\eqref{eq:9-modes} to some required order. The truncated expansion, which is a 9-variate polynomial, is then transformed by each of the 48 symmetry operations. The average of the 48 transformed polynomials is then guaranteed to be invariant under all 48 symmetries according to the rearrangement theorem.

Once we find the symmetry-determined function $E$ of the 9 modes, we project back to the 5 modes we previously started with by letting the other 4 modes $Q_1^{000}=Q_1^{\pi\pi\pi}=Q_1^{0\pi\pi}=Q_1^{\pi 0\pi}=0$. The general form of $E$ is then given by
\begin{align} \label{eq:E-5-modes}
E=\sum_{n=0}^{\infty} \sum_{j=0}^{2n} \sum_{m=0}^{\infty}
C_{njm}(Q_0,Q_3)q_0^{2n-j} q_3^j q_1^{2m}
\end{align}
where we have used the short-hand notations $Q_0=Q_0^{000}$, $Q_3=Q_3^{000}$, $q_0=Q_0^{\pi\pi\pi}$, $q_3=Q_3^{\pi\pi\pi}$, $q_1=Q_1^{\pi\pi 0}$, which we also used in the main text. The functions $C_{njm}(Q_0,Q_3)$ are Taylor expandable and have the forms
\begin{align}
C_{000}(Q_0,Q_3)&=a_0(Q_0)Q_0^2+b_0(Q_0,Q_3)Q_3^2, \label{eq:C000}\\
C_{n00}(Q_0,Q_3)&=a_n(Q_0)+b_n(Q_0,Q_3)Q_3^2,\\
C_{n10}(Q_0,Q_3)&=c_n(Q_0,Q_3)Q_3, \label{eq:Cn10}
\end{align}
where $n=1, 2, 3, \cdots$, and other $C_{njm}(Q_0,Q_3)$ functions and all lower-case functions that appear in Eqs. \eqref{eq:C000}--\eqref{eq:Cn10} \ are arbitrary Taylor-expandable functions. Eq.~\eqref{eq:E-5-modes} can be thought of as some advanced version of Eq.~\eqref{eq:E-isolated} in the main text with arbitrary constants aggregated into the Taylor coefficients of the arbitrary functions $a_n(Q_0)$, $b_n(Q_0,Q_3)$, $c_n(Q_0,Q_3)$, etc.

Now we study the strain effects, i.e. how things depend on the lattice constant $a$. Since the $Q_0$ and $Q_3$ modes are at k-point $\mathbf{F}=(0,0,0)$, they are more closely related to the value of $a$ than the other modes $q_0$, $q_3$, and $q_1$. As a simplification, we assume that $Q_0=Q_0(a)$ and $Q_3=Q_3(a)$ are smooth functions of the control parameter $a$ directly. As $a$ increases, one expects $Q_0(a)$, the overall volume expansion mode, to monotonically increase and $Q_3(a)$, the overall cubic-to-tetragonal Jahn-Teller distortion, to monotonically decrease. The other 3 modes $q_0$, $q_3$, and $q_1$ may exhibit discontinuous jumps or other non-analytic behaviors at certain critical values of $a$ and have to be kept as order parameters explicitly in the Landau energy function $E$. One may refer to the calculated structures in Fig.~\ref{fig:modes} to see that the jumps in $Q_0$ and $Q_3$ at the transitions are much smaller. The energy function $E$ is thus simplified to
\begin{align}
E=\sum_{n=0}^{\infty} \sum_{j=0}^{2n} \sum_{m=0}^{\infty}
C_{njm}(a)q_0^{2n-j} q_3^j q_1^{2m}
\end{align}
which now has only three order parameters $q_0$, $q_3$, and $q_1$. The modes $Q_0$ and $Q_3$ are treated as control parameters that are smooth functions of the lattice constant $a$ and therefore disappear from the energy function.

A further simplification can be made by noticing that in the calculated structures of LuNiO$_3$ (see Fig.~\ref{fig:modes}), the order parameters $q_0$ and $q_3$, both at the k-point $\mathbf{G}=(\pi,\pi,\pi)$, are always simultaneously nonzero, as in the ``charge-ordered'' structure, or simultaneously zero when the order is killed by a sufficiently large compressive/tensile strain. The fact that $q_0$ and $q_3$ always go ``hand in hand'' and ``die together'' suggests that we may combine them into one order parameter. This can be done by treating the ratio \ $q_3/q_0 =\lambda(a)$ as a continuous (not necessarily monotonic) function of $a$. The Landau function is now further reduced to one with only two order parameters:
\begin{align} \label{eq:E-cubic}
E=\sum_{n=0}^{\infty} \sum_{m=0}^{\infty} A_{2n,2m}(a)q_0^{2n}q_1^{2m},
\end{align}
where the coefficients
\begin{align}
A_{2n,2m}(a)=\sum_{j=0}^{2n} C_{njm}(a) \lambda^j(a)
\end{align}
are arbitrary independent continuous functions of $a$. Eq.~\eqref{eq:E-cubic} gives the general form of the symmetry-based Landau energy function of $R$NiO$_3$ without considering perovskite octahedral rotations and non-orthogonal Ni-O bond angles. The Eq.~\eqref{eq:Eq0q1} in the main text is a simplified model that suffices to explain our numerical results.

\subsection{Including octahedral rotations: a\\ perturbative approach \label{appendix:B}}

Only even powers of $q_0$ and $q_1$ enter the Landau energy function $E$ in Eq.~\eqref{eq:E-cubic}. Here, we consider the effects of the perovskite octahedral rotations in $R$NiO$_3$. We find that the rotations preserve the $q_0\leftrightarrow -q_0$ symmetry but break the $q_1\leftrightarrow -q_1$ symmetry, which allows odd powers of $q_1$ to enter the energy function $E$.

The rotational pattern in LuNiO$_3$ (and other rare-earth nickelates) is of the GdFeO$_3$ type, symbolically written as $\alpha_z^+ \beta_x^- \beta_y^-$. The meaning of the symbol is already explained in detail in the main text. We want to construct an energy $E$ as a function of the rotational angles $\alpha_z^+$, $\beta_x^-$, $\beta_y^-$ and the bond-length modes $q_0$ and $q_1$ using the symmetry group D$_{4h}+\mbox{translations}$, because at general values of the lattice constant $a$, the Jahn-Teller mode $Q_3(a)$ at the k-point $\mathbf{F}=(0,0,0)$ lowers the point group symmetry from O$_h$ (cubic) to D$_{4h}$ (tetragonal).

The benefits of using D$_{4h}$ instead of O$_h$ are a) coupling terms involving $Q_0$ and $Q_3$ are automatically allowed for and b) the modes $\alpha_z^+$, $\beta_x^-$, $\beta_y^-$, $q_0$ and $q_1$ already form a group-invariant subspace without needing any extensions. Since all axial vectors $\alpha_z^+$, $\beta_x^-$, $\beta_y^-$ and bond-length modes $q_0$, $q_1$ are invariant under spatial inversion I, only D$_{4h}/\{$E, I$\}=$ D$_4$, which contains 8 symmetry operations, is effective in actually transforming the 5 modes. In addition to D$_4$, the translations can generate 4 possible ways of sign change according to the k-points of the 5 modes, among which $\alpha_z^+$ and $q_1$ are at $\mathbf{C}=(\pi,\pi,0)$, and $\beta_x^-$, $\beta_y^-$ and $q_0$ are at $\mathbf{G}=(\pi,\pi,\pi)$. We therefore have totally $8\times 4=32$ symmetries to satisfy in order to construct a valid energy function $E(\alpha_z,\beta_x,\beta_y,q_0,q_1)$ that takes into account the (small) octahedral rotations.

Following again the algorithm in Appendix \ref{appendix:A} based on the rearrangement theorem of group theory, we get the general form of the symmetry-allowed Taylor expansion of the energy function
\begin{align} \label{eq:E-rotations}
E=&A\alpha_z^2+B(\beta_x^2+\beta_y^2)+C q_0^2+D q_1^2
\nonumber\\
&+F\alpha_z\beta_x\beta_y q_1+\cdots,
\end{align}
where all coefficients $A$, $B$, $C$, $D$, $F$, etc. can be arbitrary functions of $a$. This is because the $Q_0(a)$ and $Q_3(a)$ modes are functions of $a$ and are invariant under $D_{4h}$ and translations. They can therefore arbitrarily couple to any variables in Eq.~\eqref{eq:E-rotations}.

The omitted terms in Eq.~\eqref{eq:E-rotations} include other quartic terms that are products of the quadratic ones and higher-order terms. The leading-order term that breaks the $q_1\leftrightarrow-q_1$ symmetry is $F\alpha_z\beta_x\beta_y q_1=F\alpha\beta^2 q_1$, which is linear in $q_1$. This term exists even if one considers the full O$_h$ symmetry, which symmetrizes it to
\begin{align}
\nonumber
F\alpha_z\beta_x\beta_y q_1 & \rightarrow F(\alpha_z\beta_x\beta_y Q_1^{\pi\pi 0}\\
&+\alpha_x\beta_y\beta_z Q_1^{0\pi\pi}+\alpha_y\beta_z\beta_x Q_1^{\pi 0\pi}).
\end{align}
This means the octahedral rotations $\alpha_z^+\beta_x^-\beta_y^-$ break the $q_1\leftrightarrow-q_1$ symmetry even at $Q_3(a)=0$, i.e. at zero strain. However, the $q_0\leftrightarrow-q_0$ symmetry is strictly preserved order by order. Switching the sizes of the larger and smaller NiO$_6$ octahedra in the ``charge-ordered'' structure is still a symmetry of the system even in the presence of the GdFeO$_3$-type octahedral rotations.

We therefore add the leading order symmetry-breaking term $F\alpha_z\beta_x\beta_y q_1$ to the original Landau function $E$ in Eq.~\eqref{eq:E-cubic} as a perturbation to get the symmetry right. The new Landau function is given by
\begin{align} \label{eq:E-general}
E=\sum_{n=0}^{\infty} \sum_{m=0}^{\infty} A_{2n,2m}(a) q_0^{2n} q_1^{2m} + F(a) \alpha \beta^2 q_1.
\end{align}\\
The added term should be small because $\alpha\beta^2\ll 1$ for small rotations $\alpha$ and $\beta$. It should therefore be ineffective unless the even-power coefficients $A_{2n,2m}(a)$ make $q_1=0$ unstable or nearly unstable. The Eq.~\eqref{eq:Efinal} in the main text is a simplified model of the general Eq. \eqref{eq:E-general} here.

Aside from octahedral rotations, non-orthogonal Ni-O bond angles can also break the $q_1\leftrightarrow-q_1$ symmetry if the Ni-O bond that is approximately along the $z$ direction forms different angles with the $x$ and $y$ bonds. The leading-order symmetry-breaking term should be also small and linear in $q_1$, and can therefore be addressed in the same footing as octahedral rotations in our general model system given by Eq.~\eqref{eq:E-general}.
\end{document}